\documentclass[aps,pre,reprint,superscriptaddress]{revtex4-2}
\usepackage{amsmath,amssymb,graphicx,siunitx,xcolor,hyperref,physics-patch,booktabs,braket,bm,derivative}
\hypersetup{
colorlinks,
linkcolor={blue!50!black},
citecolor={blue!50!black},
urlcolor={blue!80!black}}
\DeclareMathOperator*{\med}{median}

\begin{document}
\title{Critical fluctuations of elastic moduli in jammed solids}
\author{Kumpei Shiraishi}
\affiliation{SANKEN, University of Osaka, 8-1 Mihogaoka, Ibaraki, Osaka 567-0047, Japan}
\author{Hideyuki Mizuno}
\affiliation{Graduate School of Arts and Sciences, University of Tokyo, 3-8-1 Komaba, Meguro, Tokyo 153-8902, Japan}
\date{\today}
\begin{abstract}
We investigate sample-to-sample fluctuations of the shear modulus in ensembles of particle packings near the jamming transition.
Unlike the average modulus, which exhibits distinct scaling behaviours depending on the interparticle potential, the fluctuations obey a critical exponent that is independent of the potential.
Furthermore, this scaling behaviour has been confirmed in two-dimensional packings, indicating that it holds regardless of spatial dimension.
Using this scaling law, we discuss the relationship predicted by heterogeneous-elasticity theory between elastic-modulus fluctuations and the Rayleigh scattering of sound waves across different pressures.
Our numerical results provide a useful foundation for developing a unified theoretical description of the jamming critical phenomenon.
\end{abstract}
\maketitle

\section{Introduction}
When particles are compressed, the system jams and acquires rigidity at a certain packing fraction.
The phenomenon, called the jamming transition, is ubiquitous in diverse materials such as foams, colloidal suspensions, pastes, and granular materials, and forms a basis for understanding properties of disordered systems, spanning from structural glasses to biological systems~\cite{Liu_2010,van_Hecke_2010}.
Over the past 30 years, various studies, including theoretical~\cite{Wyart_EPL_2005,Wyart_PRE_2005}, numerical~\cite{O_Hern_2002,O_Hern_2003} and experimental~\cite{Mason1995,Majmudar_2007} approaches, have characterised basic properties of this phenomenon.
The most central observation is made in critical phenomena; various physical quantities exhibit power-law scaling near the transition~\cite{O_Hern_2002,O_Hern_2003}.
This critical behaviour manifests in elastic moduli~\cite{O_Hern_2003,Ellenbroek_2006}, vibrations~\cite{Silbert_2005,Mizuno_2017}, transport properties~\cite{Xu_2009,Vitelli_2010,Mizuno_2018}, and rheology~\cite{Olsson_2007}.
Furthermore, at the transition point, the number of contacts exactly matches the number of degrees of freedom in the system, a state known as isostaticity~\cite{O_Hern_2003}.
It is now well established that the jamming scaling law is controlled by the contact number excess to the isostatic value~\cite{Wyart_EPL_2005,Wyart_PRE_2005}.

One of the central challenges in statistical physics is the understanding of phase transitions and critical phenomena~\cite{Stanley1971,Binney1992,Goldenfeld1992,Nishimori2010Elements}.
In this context, theoretical tools such as mean-field theory and, more importantly, the renormalization group have been developed, and it has become clear that fluctuations play a crucial role in critical behaviour.
In particular, as a system approaches a critical point, both the fluctuations and the associated correlation length scales diverge, leading to various anomalies at the critical point.
In the field of jamming transition, some works have followed this line of research, including scaling collapse and finite-size scaling~\cite{Ellenbroek_2006,Goodrich_2012,Goodrich_2014}, diverging length scales~\cite{Silbert_2005,Wyart_EPL_2005,Lerner_2014,Yan_2016}, and Widom scaling hypothesis~\cite{Goodrich_2016}.
Thus, understanding how fluctuations in physical quantities behave near the jamming critical point remains an important issue.
Despite its relevance, the study of fluctuations in jamming systems is still limited, and further development is anticipated~\cite{Hexner_2018,Hexner_2019,Ikeda2023JCP,Giannini2024}.

Beyond their relevance in statistical physics, fluctuations in elastic moduli play an essential role in the theoretical description of the anomalous properties of amorphous solids.
Among the various theoretical frameworks~\cite{Gurevich_2003,Franz_2015,Damart_2017,Baggioli2022,PhysRevLett.130.236101,Vogel2025SelfConsistent}, the heterogeneous-elasticity theory (HET)~\cite{SchirmacherRuocco,Schirmacher_2006,Schirmacher_2007,Marruzzo_2013,Schirmacher_2014} is one of the most well-established approaches for describing the anomalous behaviour of amorphous materials such as jammed solids.
HET postulates that the local shear modulus exhibits spatial fluctuations throughout the system, and extensive numerical studies have indeed confirmed the existence of such elastic heterogeneity in amorphous systems~\cite{Tsamados_2009,Mizuno_PRE_2013,Mizuno_PRL_2016}.
Within this theoretical framework, the so-called disorder parameter $\gamma$, which quantifies the variance of local shear moduli, acts as a key control parameter~\cite{Schirmacher_2006}.
In the standard formulation of HET, the distribution of local shear moduli is assumed to be Gaussian, and its mean value and standard deviation serve as inputs for computing macroscopic observables.
This formulation successfully accounts for a variety of universal features of amorphous solids.
For example, the low-frequency vibrational modes in excess over the Debye prediction (the boson peak) can be captured within HET, with both their characteristic frequency and intensity expressed as functions of $\gamma$.
Likewise, the theory predicts a Rayleigh-type frequency dependence of the acoustic attenuation, $\Gamma \sim \Omega^{d+1}$ ($\Gamma$ is the attenuation coefficient and $\Omega$ is the propagation frequency), which is also observed in experiments on glasses~\cite{Baldi_2010,Baldi_2014}.
The prefactor of this relation, corresponding to the strength of Rayleigh scattering, is again determined by $\gamma$, linking vibrational and transport anomalies to the underlying elastic heterogeneity.
These theoretical predictions, spanning both vibrational spectra~\cite{Mizuno_PRB_2016,Mizuno_2018,Shiraishi2023pin2D} and acoustic attenuation~\cite{Mizuno_Mossa_Barrat_2014,Mizuno_2018,Wang_sound_sttenuation_2019}, have been extensively tested through numerical simulations.

Another major theoretical framework developed to describe and predict the properties of amorphous solids is the effective medium theory (EMT), which is based on spring-network models~\cite{DeGiuli_2014}.
While HET treats the spatial fluctuations of local shear moduli as the sole mesoscale input, EMT instead relies on microscopic parameters, namely the mean coordination number $z$ and the contact strain $e$, often referred to as the pre-stress.
Despite this difference in the scale of description, EMT successfully reproduces the same key features of vibrational and scattering phenomena as HET, including the emergence of the boson peak and the Rayleigh-type acoustic attenuation.
Given that both theories capture similar physical phenomena from different levels of description, it is natural to expect that they are connected via an appropriate coarse-graining procedure.
However, the precise nature of this connection remains elusive.
One major reason for this persistent gap is the limited numerical characterisation of fluctuations of elastic modulus.
From this perspective, a detailed characterisation of local shear modulus fluctuations is of particular theoretical significance, particularly for bridging the gap between these two frameworks.

In this work, we characterise the sample-to-sample fluctuations of elastic moduli in ensembles of jammed packings by means of numerical simulation.
These fluctuations are believed to encode information equivalent to the disorder parameter in HET~\cite{Kapteijns_2021,Mahajan_2021}.
We focus on two prototypical models of jammed amorphous solids: harmonic spheres and Hertzian spheres.
Our results reveal that, while the average elastic modulus exhibits a model-dependent scaling behaviour, its fluctuations diverge in a manner independent of the interaction potential.
This critical scaling is also observed in two-dimensional packings.
Finally, we discuss the theoretical implications of these findings, with particular emphasis on their relevance to the HET prediction on acoustic attenuation in amorphous solids.

\section{Methods}
\label{sec:Methods}
\subsection{System description}
We consider randomly jammed packings in a three-dimensional box with periodic boundary conditions.
Particles interact with a finite-range purely repulsive potential~\cite{O_Hern_2003}
\begin{align}
v(r_{ij}) = \frac{\epsilon}{\alpha} \pqty{1 - \frac{r_{ij}}{\sigma_{ij}}}^\alpha H\pqty{\sigma_{ij} - r_{ij}}
\end{align}
where $r_{ij}$ is the distance between particles $i$ and $j$, $\sigma_{ij} = (\sigma_i + \sigma_j)/2$ is the sum of radii of two particles, and $H\pqty{x}$ is the Heaviside step function.
Two models of jammed particles are considered, harmonic spheres ($\alpha = 2$) and Hertzian spheres ($\alpha = 5/2$).
For both models, we use 50:50 binary mixtures of particles whose diameters are $\sigma$ and $1.4\sigma$ with identical mass $m$.
The number of particles in the system is denoted as $N$.
Masses, lengths, and energies are measured in units of $m$, $\sigma$, and $\epsilon$, respectively.
The number density is expressed as $\rho = N/L^d$, where $d = 3$ is the spatial dimension.

\subsection{Numerical generation of jammed packings}
We start the preparation of packings by minimising the energy of random configurations at a sufficiently high packing fraction $\varphi_\mathrm{init} = 1.0$ with the FIRE algorithm~\cite{Guenole_2020}.
The algorithm is terminated when the condition $\max_i F_i < \num{e-12}$ is reached.
Then, we perform the global compression/decompression to generate packings at a given target pressure $p$~\cite{GoodrichThesis}.
For harmonic spheres, pressures range over $p = \num{e-2}, \num{e-3}, \dots, \num{e-6}$ for all system sizes.
For Hertzian spheres, the highest pressure is $p = \num{e-2}$ for all system sizes, and the lowest pressure is $p = \num{e-8}$ for $N = \num{2000}$.
For the case of $N = \num{32000}$, the lowest pressure is $p = \num{e-7}$.
A packing ensemble prepared with the compression/decompression-only protocols may contain samples that have finite pressure but are unstable to shear~\cite{Dagois_Bohy_2012}.
We discard such samples with a negative shear modulus when constructing the ensembles.
The sizes of ensembles used in this study are given in Appendix~\ref{sec:appendix}.
After packings are generated, we recursively remove rattler particles whose number of contacts $N_c$ satisfies $N_c \leq d$~\cite{Charbonneau_PRL_2015}.

\subsection{Calculation of stressed and unstressed shear moduli}
Here, we recap the linear response formulation of shear moduli used in this study~\cite{Maloney_2004,Maloney_2006,Lemaitre_2006}.
Because the system is isotropic, we measure the elastic response exclusively for a shear strain in the $x$ direction with a strain gradient in the $y$ direction.

The elastic modulus is decomposed into two components, the affine and non-affine modulus: $G = G_A - G_N$~\cite{Alexander_1998,Ashcroft_Mermin}.
The affine modulus (Born--Huang term) is the second derivative of total potential energy $E$, defined as
\begin{align}
E = \sum_{\Braket{ij}}v(r_{ij}),
\end{align}
with respect to the affine strain, evaluated as~\cite{Maloney_2004,Maloney_2006,Lemaitre_2006,Mizuno_Saitoh_Silbert_2016,Lutsko_1989,Barron_1965}
\begin{align}
G_A = \frac{1}{V}\sum_{\Braket{ij}} \bqty{\pqty{k_{ij} + \frac{f_{ij}}{r_{ij}}} \frac{x_{ij}^2 y_{ij}^2}{r_{ij}^2} - f_{ij}\frac{r_{ij}}{d}},
\label{eq:affine_modulus}
\end{align}
where $V$ is the volume of the simulation box, $f_{ij} = -v^\prime\pqty{r_{ij}}$ and $k_{ij} = v^{\prime\prime}\pqty{r_{ij}}$ are the first and second derivatives of the potential $v\pqty{r_{ij}}$ with respect to $r_{ij}$.
The distance $r_{ij}$ is given by the norm of the displacement vector $\bm{r}_{ij} = \pqty{x_{ij}, y_{ij}, z_{ij}}$.
The sum is taken for all bonds $\Braket{ij}$ in the system.
We note that the second term in the summation of Eq.~\eqref{eq:affine_modulus} is a correction term for the finite pressure~\cite{Wittmer2013EPJE,Wittmer2013JCP}.

The second term $G_N$ is due to non-affine relaxation.
This term is evaluated with the Hessian matrix~\cite{Maloney_2004,Maloney_2006,Lemaitre_2006}
\begin{align}
G_N = \frac{1}{V} \Xi \cdot \mathcal{H}^{-1} \cdot \Xi,
\end{align}
where the $i$-th block of the affine force $\Xi$ ($3N$-dimensional vector) is provided as
\begin{align}
\Xi_i = \sum_{j \neq i} \pqty{k_{ij} + \frac{f_{ij}}{r_{ij}}} \frac{x_{ij}y_{ij}}{r_{ij}} \frac{\bm{r}_{ij}}{r_{ij}},
\end{align}
and the Hessian matrix (size $3N \times 3N$) is defined as
\begin{align}
\mathcal{H} = \bqty{\pdv{E}{\bm{r}_i,\bm{r}_j}}_{i,j = 1,2,\dots,N},
\end{align}
where $\bm{r}_i = \pqty{x_i, y_i, z_i}$ is the coordinate of particle $i$.
In amorphous solids, affine deformation causes an additional non-affine displacement field $\delta\bm{u} = \mathcal{H}^{-1} \cdot \Xi$~\cite{Alexander_1998,Maloney_2004,Maloney_2006,Lemaitre_2006,Zaccone_2011}.
This displacement field $\delta\bm{u}$ ($3N$-dimensional vector) is usually obtained though the decomposition of $\Xi$ with the normal modes of $\mathcal{H}$~\cite{Maloney_2004,Maloney_2006,Lemaitre_2006}.
However, this method is computationally hard when the size of matrix $\mathcal{H}$ becomes large.
Therefore, we obtain the non-affine displacements $\delta\bm{u}$ via the FIRE minimisation of a cost function
\begin{align}
L = \frac{1}{2} \delta\bm{u} \cdot \mathcal{H} \cdot \delta\bm{u} + \delta\bm{u} \cdot \Xi
\label{eq:costfunc}
\end{align}
instead of performing the direct diagonalization of the Hessian matrix~\cite{Hara2023}.
In this minimisation, the termination condition is set to $\max_i \abs{-\partial L / \partial\delta\bm{u}_i} < \num{e-8}$.
This iterative method yields displacements that are identical to the direct solution of the linear equation $\mathcal{H}\cdot\delta\bm{u} = \Xi$ within numerical tolerance.

In the harmonic limit, the energy change $\Delta E$ caused by relative displacements $\Bqty{\bm{u}_i}$ from the mechanically equilibrated position $\Bqty{\bm{r}_i}$ takes the form~\cite{Wyart_EPL_2005,Wyart_PRE_2005,Ellenbroek_2006,Ellenbroek_van-Hecke_2009,Mizuno_Saitoh_Silbert_2016}
\begin{align}
\Delta E = \frac{1}{2} \sum_{\Braket{ij}} \pqty{k_{ij} u_{ij,\parallel}^2 - \frac{f_{ij}}{r_{ij}} u_{ij,\perp}^2},
\label{eq:harmonic_energy}
\end{align}
where $u_{ij,\parallel}$ and $u_{ij,\perp}$ denote the components of the displacement parallel and perpendicular to $\bm{r}_{ij}$, respectively.
This may be written compactly using the Hessian matrix: $\Delta E = \bm{u}\cdot\mathcal{H}\cdot\bm{u}$.
Because the contact force $f_{ij} = -v^\prime\pqty{r_{ij}}$ is always repulsive in the jammed packings, the original state is called the stressed system.
On the other hand, we have also studied the shear modulus of the unstressed system~\cite{Wyart_PRE_2005}, where we remove $f_{ij}$ in the Hessian matrix while maintaining $k_{ij}$.
This case corresponds to the system where we replace the interactions between particles with relaxed (unstretched) springs of each stiffness $k_{ij}$.
In practice, we omit the first derivatives in the Hessian matrix in the cost function \eqref{eq:costfunc} when calculating the shear modulus of the unstressed system.
We denote the shear modulus of the stressed system by $G_1$ and that of the unstressed system by $G_0$.

\section{Behaviour of average shear modulus}
\begin{figure}
\centering
\includegraphics[width=\linewidth]{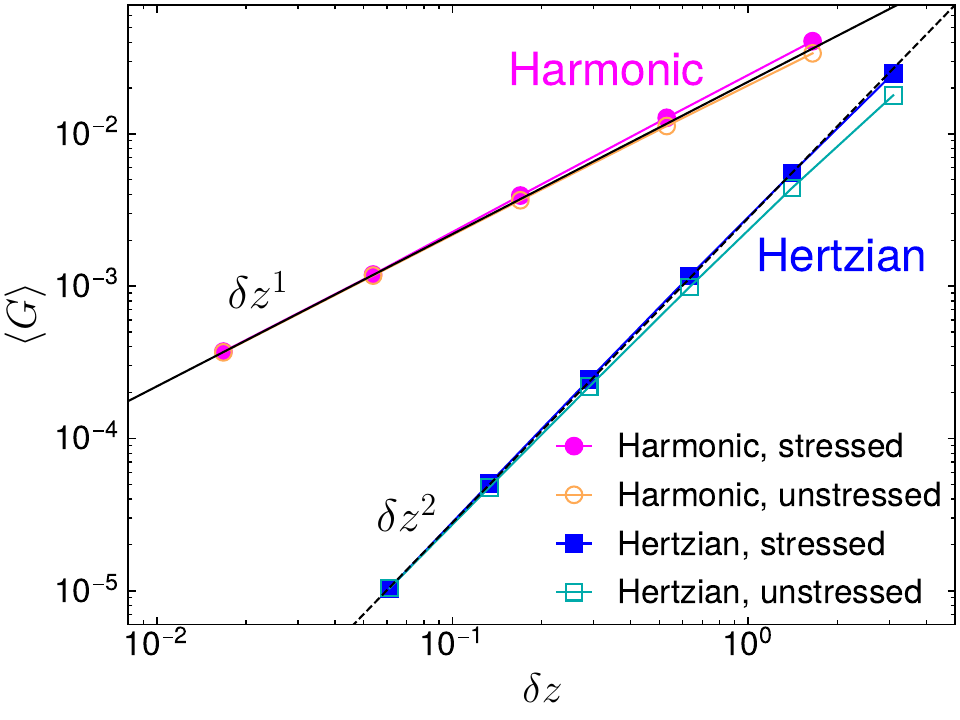}
\caption{Critical scaling of the average shear moduli $\Braket{G_1}$ and $\Braket{G_0}/2$ for harmonic and Hertzian packings with $N = \num{32000}$. The solid and dashed lines represent $\Braket{G} \sim \delta z$ and $\Braket{G} \sim \delta z^2$, respectively.}
\label{fig:G_mean}
\end{figure}

First, we present the critical behaviour of the ensemble-averaged shear modulus.
As reported in numerous studies~\cite{O_Hern_2003,Ellenbroek_2006,Makse_1999,Wang_2021}, the shear modulus exhibits the following critical scaling with pressure $p$:
\begin{align}
G \sim p^{\pqty{\alpha-3/2} / \pqty{\alpha-1}}.
\end{align}
In addition, the excess contact number $\delta z = z - z_\mathrm{iso}$, which quantifies the distance from the isostatic point at the onset of jamming~\cite{O_Hern_2003,Wyart_PRE_2005}, is known to scale as~\cite{O_Hern_2003}:
\begin{align}
\delta z \sim p^{1/\pqty{2\alpha-2}}.
\end{align}
Here, $z_\mathrm{iso} = 2d\pqty{1 - 1/N}$ is the isostatic contact number that satisfies the Maxwell criterion, and the $1/N$ correction arises from the global zero-frequency translational modes of the finite-size systems~\cite{Goodrich_2012}.
Combining the two scaling relations yields $G \sim \delta z$ for harmonic spheres ($\alpha = 2$), and $G \sim \delta z^2$ for Hertzian spheres ($\alpha = 5/2$).
As clearly shown in Fig.~\ref{fig:G_mean}, our numerical data are consistent with these predictions:
harmonic spheres exhibit the scaling $\Braket{G_1} \sim \delta z$, while Hertzian spheres obey $\Braket{G_1} \sim \delta z^2$, where $\Braket{\cdot}$ denotes the ensemble average.

We now turn to the unstressed modulus $G_0$.
Similar to the stressed modulus $G_1$, the unstressed modulus $G_0$ follows the same scaling laws for harmonic and Hertzian spheres, as shown in Fig.~\ref{fig:G_mean}.
In addition, according to the EMT for spring networks, the stressed shear modulus $G_1$ is expected to be twice the unstressed modulus $G_0$ near the jamming point~\cite{DeGiuli_2014}.
To test this prediction, we plot $\Braket{G_0}/2$ for both types of packings in Fig.~\ref{fig:G_mean}.
Near the jamming transition, $\Braket{G_0}/2$ collapses onto $\Braket{G_1}$, indicating good quantitative agreement with EMT.
However, further from the transition, $\Braket{G_1}$ exceeds $\Braket{G_0}/2$, suggesting that the pre-stress $e$ at higher pressure deviates from its critical value $e_c$~\cite{DeGiuli_2014}.
Overall, Fig.~\ref{fig:G_mean} clearly demonstrates that both $\Braket{G_0}$ and $\Braket{G_1}$ obey the same scaling law within each model, while the scaling exponent itself depends on the interaction potential.

\section{Divergence of critical fluctuations}
In this section, we examine the sample-to-sample probability distributions within the ensemble used to compute the average quantities discussed above.
The discussion is divided into two subsections, focusing respectively on the contact number and the shear modulus.
In each case, we analyse the probability distribution of the observable of interest and quantify its fluctuations.

\subsection{Excess contact number}
\begin{figure}
\centering
\includegraphics[width=\linewidth]{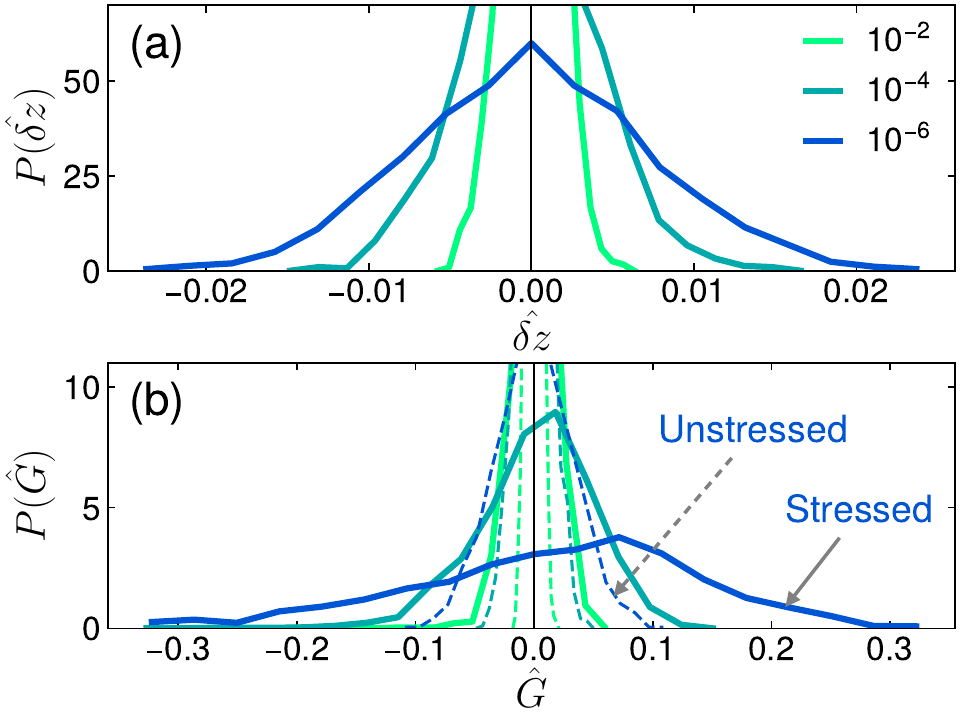}
\caption{Probability distributions of (a) the excess contact number $\delta z$ and (b) the shear modulus $G$ for the ensemble of Hertzian packings ($N = \num{32000}$) at various pressures.
To enable comparison between different pressures, each quantity $X$ is rescaled as $\hat{X} = X/\Braket{X} - 1$.
In panel (b), the distributions of the stressed modulus $G_1$ are shown as solid curves, whereas those of the unstressed modulus $G_0$ appear as dashed curves.}
\label{fig:PDF_z_G_Hertzian}
\end{figure}

We begin by presenting the sample-to-sample probability distribution of the excess contact number $\delta z$.
Figure~\ref{fig:PDF_z_G_Hertzian} (a) shows the distribution $P(\hat{\delta z})$ computed from the ensemble of Hertzian packings with $N = \num{32000}$.
To facilitate comparison across different pressures, we rescale the variable as $\hat{\delta z} = \delta z/\Braket{\delta z} - 1$.
As illustrated in Fig.~\ref{fig:PDF_z_G_Hertzian} (a), the distribution is symmetric and approximately Gaussian in shape.
As the pressure decreases and the system approaches the jamming transition, the distribution broadens.

We next quantify the pressure dependence of these fluctuations across the ensemble.
To do so, we adopt a disorder quantifier for a generic observable $X$, defined as~\cite{Giannini2024}
\begin{align}
\chi_X = \frac{\sqrt{N\Braket{\pqty{X - \Braket{X}}^2}}}{\Braket{X}} = \frac{\sqrt{N}\sigma_X}{\Braket{X}},
\end{align}
where $\sigma_X$ is the standard deviation.
We use this measure to characterise the fluctuations in the excess contact number ($X = \delta z$).

\begin{figure}
\centering
\includegraphics[width=\linewidth]{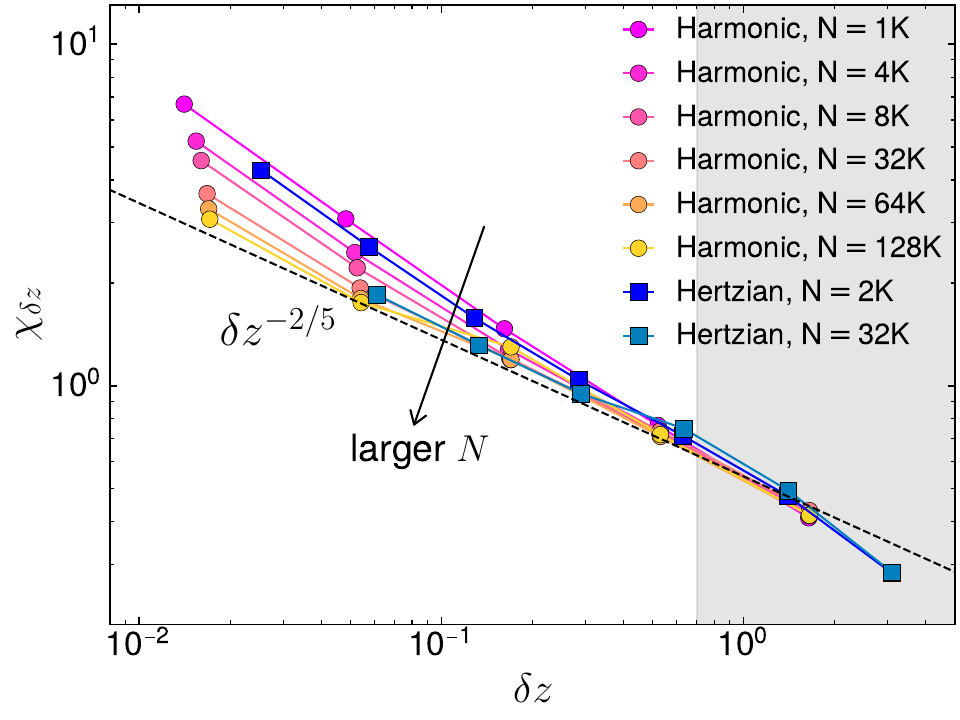}
\caption{Critical fluctuations of the excess contact number $\delta z$.
The dashed line represents the scaling $\chi_{\delta z} \sim \delta z^{-2/5}$.
The shaded region marks the glassy regime, $\delta z \gtrsim \num{7e-1}$, where this scaling behaviour is no longer expected to hold~\cite{Giannini2024}.}
\label{fig:chi_dz}
\end{figure}

Figure~\ref{fig:chi_dz} displays $\chi_{\delta z}$ for both interaction types across various system sizes.
Harmonic spheres exhibit scaling behaviour $\chi_{\delta z} \sim \delta z^{-2/5}$, consistent with the findings of Giannini and coworkers~\cite{Giannini2024}, albeit subject to finite-size effects.
On the other hand, contact-number fluctuations $\chi_{\delta z}$ of the Hertzian spheres follow a similar scaling behaviour with $\delta z$.
Giannini and coworkers identified distinct regimes within the jammed phase based on the behaviour of $\chi_{\delta z}$.
They reported that $\chi_{\delta z}$ becomes independent of system size at high pressures, where the scaling becomes steeper, with a more negative exponent.
According to their analysis, this crossover occurs at $\delta z \approx \num{7e-1}$, and they referred to the corresponding high-$\delta z$ region as the glassy regime ($\delta z \gtrsim \num{7e-1}$).
Our results for both harmonic and Hertzian spheres are consistent with these observations, although we do not explore this aspect in further detail here.

\subsection{Shear modulus}
We now turn to the sample-to-sample fluctuations of the shear modulus.
Figure~\ref{fig:PDF_z_G_Hertzian} (b) displays the probability distributions of the shear moduli computed from the ensemble of Hertzian packings.
Both the stressed modulus $G_1$ and the unstressed modulus $G_0$ are shown, represented by solid and dashed curves, respectively.

Compared with the excess contact number $\delta z$ (Fig.~\ref{fig:PDF_z_G_Hertzian} (a)), both moduli exhibit broader distributions at the same system size.
As the pressure is lowered and the system approaches the jamming transition, these distributions become broader, mirroring the trend observed in $P(\delta z)$.
The distinction between $G_1$ and $G_0$ is also evident in Fig.~\ref{fig:PDF_z_G_Hertzian} (b).
At a given pressure, $G_0$ displays a narrower distribution than $G_1$.
Although both distributions broaden as pressure decreases, $G_0$ consistently remains narrower than $G_1$, yet both are wider than the distribution of $\delta z$.

The shapes of the distributions also differ notably.
At high pressure ($p = \num{e-2}$), both $G_0$ and $G_1$ exhibit symmetric distributions.
However, at lower pressures (e.g., $p = \num{e-4}$), the peak of $P(\hat{G}_1)$ shifts towards positive values ($\hat{G}_1 > 0$), and the distribution becomes skewed.
This asymmetry becomes more pronounced as the system approaches the jamming point ($p = \num{e-6}$).
By contrast, $P(\hat{G}_0)$ retains a symmetric, Gaussian-like shape throughout this pressure range.

Such non-Gaussian behaviour in the distribution of $G_1$ has been previously reported~\cite{Kapteijns_2021}, and the skewness is attributed to spatially localised vibrational modes at low frequencies, which are suppressed in the unstressed system~\cite{Mizuno_2017}.
Moreover, outliers are found only in the distribution of the stressed modulus.
To quantify this, we examine the minimum and maximum values of $\hat{G}_0$ and $\hat{G}_1$ within the ensemble.
While these rare samples are not visible in Fig.~\ref{fig:PDF_z_G_Hertzian} (b) due to limited statistics, we observe that $\lvert \min \hat{G}_1 \rvert / \max \hat{G}_1 \approx 2.5$ to $4$, indicating the presence of samples with anomalously small $G_1$.
Consequently, a Shapiro--Wilk test applied to the raw data returns $p$-values close to zero, confirming strong deviation from normality.
In contrast, for the unstressed modulus $G_0$, we find $\lvert \min \hat{G}_0 \rvert / \max \hat{G}_0 \approx 1$, indicating an absence of outliers.
The Shapiro–Wilk test on $G_0$ returns $p$-values of approximately 0.8 or larger, even without removing any outlier data, suggesting consistency with a Gaussian distribution.

We now quantify the sample-to-sample fluctuations in the shear moduli.
As noted above, the distribution of $G_1$ is markedly different from Gaussian and includes significant outliers, making the standard definition of the fluctuation $\chi_G$ problematic.
To address this, a more robust, median-based measure has been proposed~\cite{Kapteijns2021}:
\begin{align}
\chi_{X,\mathrm{med}} = \frac{\sqrt{N\med\bqty{\pqty{X - \Braket{X}}^2}}}{\Braket{X}}
\end{align}
which avoids the need for explicit outlier-exclusion protocols while remaining statistically meaningful.
We adopt this definition for both stressed and unstressed moduli ($X = G_0, G_1$).

\begin{figure}
\centering
\includegraphics[width=\linewidth]{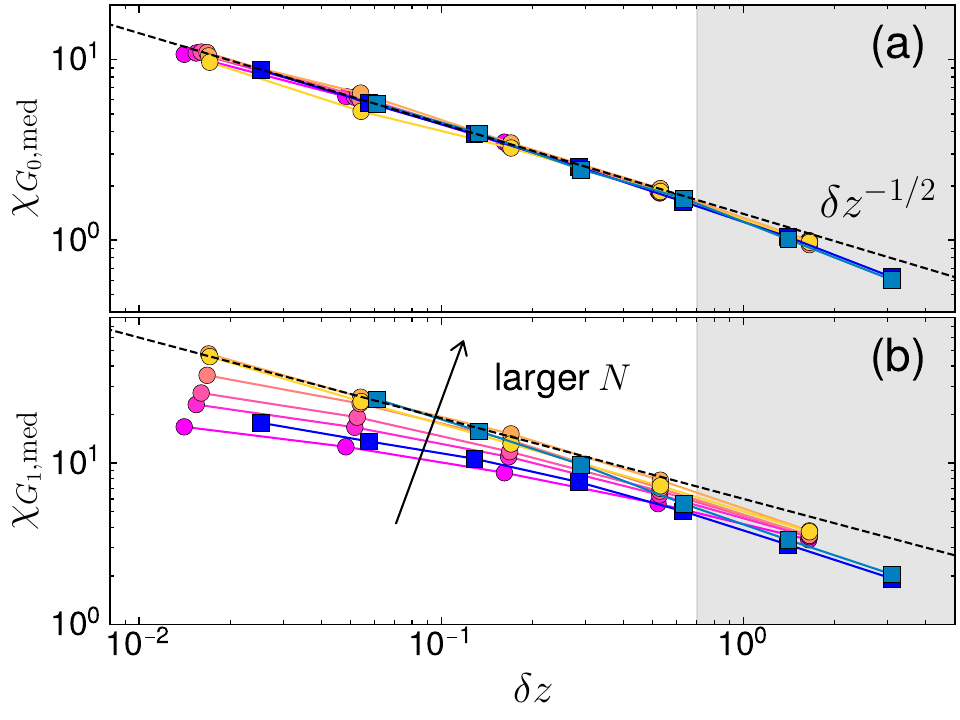}
\caption{Critical fluctuations of the shear moduli.
The unstressed modulus $G_0$ and stressed modulus $G_1$ are shown in panels (a) and (b), respectively.
The dashed lines indicate the scaling $\chi_G \sim \delta z^{-1/2}$.
The legend is identical to that in Fig.~\ref{fig:chi_dz}.
The shaded region denotes the glassy regime~\cite{Giannini2024}.}
\label{fig:chi_G}
\end{figure}

Figure~\ref{fig:chi_G} shows the fluctuation of the shear modulus as a function of $\delta z$ for both moduli.
For the unstressed modulus $G_0$, the fluctuations exhibit a clear scaling relation: $\chi_G \sim \delta z^{-1/2}$.
Finite-size effects appear negligible in this case, consistent with the narrower distribution of $G_0$ compared to $G_1$ (Fig.~\ref{fig:PDF_z_G_Hertzian} (b)).
For the stressed modulus $G_1$, the fluctuations approach the same scaling $\chi_G \sim \delta z^{-1/2}$ as the system size increases.
At higher pressures (i.e., larger $\delta z$), the scaling breaks down, consistent with the identification of the glassy regime at $\delta z \gtrsim \num{7e-1}$ by Giannini and coworkers~\cite{Giannini2024}.
An important conclusion from Fig.~\ref{fig:chi_G} is that, in contrast with the average shear modulus (Fig.~\ref{fig:G_mean}), the fluctuation $\chi_G$ appears independent of the interaction potential exponent $\alpha$.
For harmonic spheres, the scaling exponent is in agreement with previous findings~\cite{Goodrich_2014,Giannini2024}.

\section{Results on two-dimensional packings}
Finally, we perform a similar analysis on two-dimensional packings.
We use ensembles of binary disks interacting via harmonic and Hertzian potentials described in Sec.~\ref{sec:Methods}~\cite{O_Hern_2003}.
As in three dimensions, the ensemble sizes used for this calculation are listed in Appendix~\ref{sec:appendix}.
Figure~\ref{fig:chi2D} shows the jamming scaling of fluctuations of the excess contact number (panel (a)), the unstressed shear modulus (panel (b)), and the stressed shear modulus (panel (c)).
Analysis of the excess contact number $\delta z$ reveals that the contact-number fluctuation $\chi_{\delta z}$ in two dimensions follows a different critical exponent than in three dimensions: $\chi_{\delta z} \sim \delta z^{-3/5}$ (Fig.~\ref{fig:chi2D} (a)), based on our numerical data.
The dimensional difference in contact-number fluctuations was also reported in the literature~\cite{Hexner_2019} using a slightly different metric.
By contrast, our analysis of shear moduli demonstrates the shear-moduli fluctuations in two dimensions diverge in the same manner as in three-dimensional systems, $\chi_G \sim \delta z^{-1/2}$ (Fig.~\ref{fig:chi2D} (b), (c)).
This observation is consistent with previous results for harmonic packings~\cite{Goodrich_2014}, which suggested $\sigma_{G_1}/\Braket{G_1} \sim \pqty{pN^2}^{-1/4} \sim \delta z^{-1/2}$ in both two and three dimensions.
Our results therefore indicate that shear-modulus fluctuations are dimensionally independent, whereas contact-number fluctuations are not.
Notably, the robustness with respect to the interaction potential holds for fluctuations of both quantities.

\begin{figure}
\centering
\includegraphics[width=\linewidth]{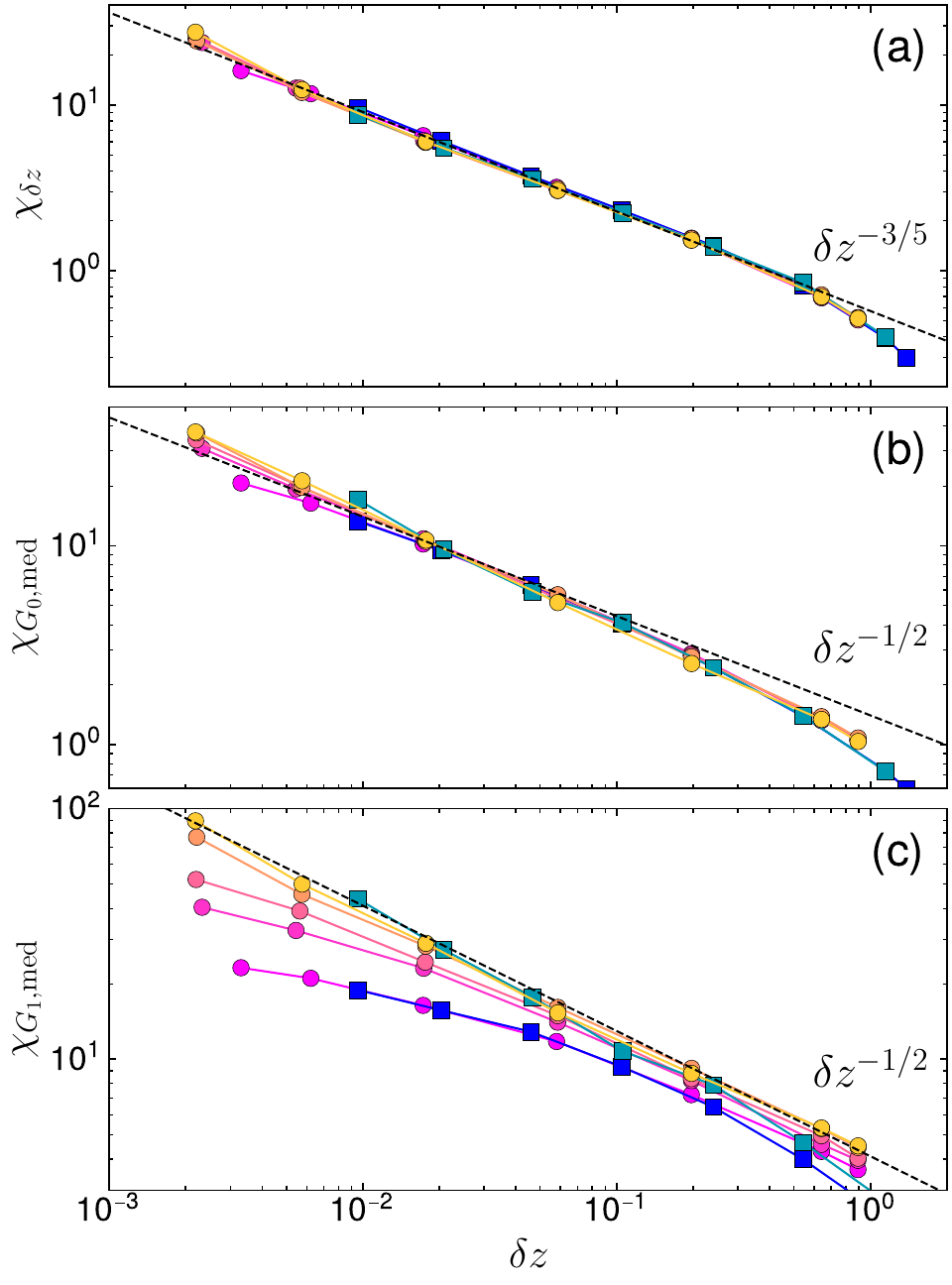}
\caption{Critical fluctuations in two-dimensional disk packings of harmonic and Hertzian interactions.
Panels (a), (b), and (c) display data of the excess contact number, the unstressed modulus, and the stressed modulus, respectively.
In panel (a), the dashed line in black represents $\chi_{\delta z} \sim \delta z^{-3/5}$.
Dashed lines in panels (b) and (c) represent $\chi_G \sim \delta z^{-1/2}$.
The legend is same with the case of three dimensions.}
\label{fig:chi2D}
\end{figure}

\section{Conclusion and discussions}
In this paper, we have investigated the sample-to-sample fluctuations of the shear modulus and contact number in two models of jammed packings.
We have found that, while the average values exhibit critical exponents that depend on the interaction potential, the critical exponents of fluctuations display potential-independent values with respect to $\delta z$.
Specifically, the fluctuation indicators scale as $\chi_{\delta z} \sim \delta z^{-2/5}$ for the excess contact number and $\chi_G \sim \delta z^{-1/2}$ for the shear modulus in three dimensions.
The exponent of the shear modulus is identical for both stressed and unstressed systems.
While the exponent of excess contact number is dimension-dependent, the exponent of shear moduli is identical for two- and three-dimensional packings.

As stated in the Introduction, the divergence of fluctuations implies the divergence of associated correlation lengths.
Our results indicate that the exponent governing the length scale associated with contact-number fluctuations depends on spatial dimension, whereas the length scale associated with elastic-modulus fluctuations exhibits the same scaling behaviour in both two and three dimensions.
The exponent controlling the divergence of elastic-modulus fluctuations coincides with that of the correlation length $l_c \sim \delta z^{-1/2}$, which characterises low-frequency vibrational modes and diverges at the jamming transition~\cite{Silbert_2005}.
Crucially, the shear modulus of jammed packing is significantly influenced by non-affine responses to the deformation, which are known to be spatially organised over $l_c$ under global shear~\cite{During_2013}.
Just as the four-point susceptibility reflects the growth of domain size of dynamic heterogeneity in glass-forming liquids, the divergence of $\chi_G$ manifests the increasing correlation length of the non-affine displacement fields.
As the system approaches the jamming transition, these displacements become increasingly correlated over larger distances, leading to greater variability in the shear modulus between different samples.
This length scale $l_c$ related to elastic-modulus fluctuations also manifests itself in vibrational modes~\cite{Silbert_2005,Yan_2016,Shimada_Mizuno_Wyart_Ikeda_2018} and responses to local perturbation~\cite{Lerner_2014}, and these behaviours can also be captured within EMT~\cite{Wyart_EPL_2010,DeGiuli_2014}.
The fact that $\chi_G \sim \delta z^{-1/2} \sim l_c$ holds across different interaction potentials and spatial dimensions strongly suggests that the elastic criticality is robustly controlled by this mesoscopic length scale, regardless of local microscopic details.
By contrast, the physical implication of the length scale associated with contact-number fluctuations remains unclear.
However, our results highlight the fact that those fluctuations obey different scaling exponents in two and three dimensions, while the average contact number follows a single exponent.
This suggests that the spatial correlation of the contact network or geometric constraints may involve different characteristic features in dimensions.

Finally, we situate our results on elastic fluctuations in the context of the HET, which provides a theoretical framework for describing sound attenuation in amorphous solids.
According to HET, the attenuation coefficient $\Gamma$ and the propagation frequency $\Omega$ are related through the Rayleigh scaling~\cite{SchirmacherRuocco}
\begin{align}
\frac{\Gamma(\Omega)}{\omega_0} \propto \gamma \pqty{\frac{\Omega}{\omega_0}}^{d+1},
\label{eq:Rayleigh}
\end{align}
where the disorder parameter $\gamma$ is determined by spatial fluctuations of the local elastic modulus.
Here, $\omega_0$ denotes the characteristic frequency associated with the elastic correlation length~\cite{Schirmacher2010,Schirmacher_2011}.
Several specific forms of $\omega_0$ have been proposed, some taking it as the boson peak frequency $\omega_\mathrm{BP}$~\cite{Mahajan_2021}, or evaluating it as $\omega_G = c_T/a_0$~\cite{Kapteijns_2021}, where $c_T$ is the sound speed of transverse wave $c_T = \sqrt{G / (m\rho)}$ and $a_0 = \rho^{-1/d}$ is the characteristic interparticle spacing.
It should be noted that HET does not provide a fully quantitative prediction of the absolute magnitude of $\Gamma$, and a system-dependent rescaling factor is generally required to achieve quantitative agreement with numerical data~\cite{Kapteijns_2021,Caroli_2019}.

Detailed numerical simulations have elucidated the jamming scaling laws governing sound attenuation~\cite{Mizuno_2018}.
In the low-frequency regime of the Rayleigh scattering, it follows
\begin{align}
\frac{\Gamma(\Omega)}{\omega^*} \propto B \pqty{\frac{\Omega}{\omega^*}}^{d+1}
\end{align}
for both two and three dimensions ($d = 2, 3$), where $B$ represents a constant of the attenuation strength.
Here, $\omega^*$ denotes the onset frequency of the plateau in the vibrational density of states, and it follows the same jamming scaling law as $\omega_\mathrm{BP}$: $\omega^* \sim \omega_\mathrm{BP} \sim \delta z$~\cite{Silbert_2005,Mizuno_2017}.
Consequently, if one adopts the HET prediction with $\omega_0 = \omega_\mathrm{BP}$, the predicted prefactor $\gamma$ becomes a constant $\gamma \sim \delta z^0$ in two and three dimensions.
By contrast, if one instead chooses $\omega_0 = \omega_G$, whose jamming scaling is $\omega_G \sim \delta z^{1/2}$, the corresponding prefactor $\gamma$ becomes dimension-dependent: $\gamma \sim \delta z^{-1}$ in two dimensions and $\gamma \sim \delta z^{-3/2}$ in three dimensions.

Recent simulations of soft-core and Lennard-Jones systems have demonstrated that this disorder parameter could be estimated from sample-to-sample fluctuations~\cite{Kapteijns_2021,Mahajan_2021,Szamel2025Perspective} by assuming the spatial correlations of local elastic moduli are short-ranged and saturate upon coarse-graining~\cite{Tsamados_2009,Mizuno_PRE_2013,Mizuno_Saitoh_Silbert_2016,Shakerpoor_2020}.
This approach also has the merit of sidestepping the direct evaluation of local elastic moduli~\cite{Mizuno_PRE_2013,Mizuno_Mossa_Barrat_2014}, and it is claimed that \emph{spatial} fluctuations of shear modulus $\gamma$ can be replaced by the \emph{sample-to-sample} fluctuations $\chi_G^2$~\cite{Kapteijns_2021,Mahajan_2021}.
Our present result of jammed packings shows that $\chi_G^2 \sim \delta z^{-1}$ regardless of the spatial dimension, and this is indeed aligned with numerical simulations of acoustic attenuation in two-dimensional packings~\cite{Mizuno_2018} if we select $\omega_G$ as $\omega_0$ in Eq.~\eqref{eq:Rayleigh} (note that the jamming scaling of the disorder parameter $\gamma$ varies on the selection of the normalisation frequency $\omega_0$).

Overall, our findings highlight that jamming scaling laws play a central role in governing elastic fluctuations and their impact on the acoustic attenuation.
Nonetheless, the numerical character of the disorder parameter employed in HET remains poorly characterised;
clarifying which fluctuations it captures and their scaling near jamming is necessary before HET can be applied to the attenuation prefactor.
To resolve these issues, systematic studies at lower pressures and larger system sizes, direct measurements of spatial correlations of local elastic moduli, and quantitative comparisons with acoustic scattering data will be crucial~\cite{Caroli_2019,Caroli_2020,Szamel2022NonAffineAttenu,Mahajan2024,FlennerSzamel2025damping,Flenner2025JPCB}.
Pursuing these directions should clarify how universal the observed scalings are and under which physical conditions the HET-based description becomes predictive.

\appendix
\section{Enumeration of the number of packings}
\label{sec:appendix}
In this appendix, we present the sizes of packing ensembles that are used in the present study.
Table~\ref{table:Nsample3D} shows the numbers of three-dimensional packings, whereas Table~\ref{table:Nsample2D} shows the numbers of two-dimensional packings.
All cases of the number of particles $N$, pressures $p$, and interaction potentials are presented in the two tables.

\begin{table*}
\centering
\caption{The number of samples of three-dimensional packings. The numbers are presented for each number of particles $N$ and each pressure $p$.}
\sisetup{table-format=6}
\begin{tabular}{SSSSSSSS}
\toprule
\text{$N$ / $p$} & \text{$10^{-2}$} & \text{$10^{-3}$} & \text{$10^{-4}$} & \text{$10^{-5}$} & \text{$10^{-6}$} & \text{$10^{-7}$} & \text{$10^{-8}$}\\
\midrule
\text{Harmonic} & & & & & & & \\
1000       & 11958            & 11825            & 11472            & 10693            & 10691            & \text{-}         & \text{-} \\
4000       & 10192            & 10157            & 10054            & 9832             & 9820             & \text{-}         & \text{-} \\
8000       & 5800             & 5796             & 5751             & 5700             & 5700             & \text{-}         & \text{-} \\
32000      & 1530             & 1530             & 1527             & 1513             & 1510             & \text{-}         & \text{-} \\
64000      & 690              & 690              & 688              & 681              & 681              & \text{-}         & \text{-} \\
128000     & 407              & 407              & 407              & 407              & 405              & \text{-}         & \text{-} \\
\midrule
\text{Hertzian} & & & & & & & \\
2000       & 8047             & 8044             & 8014             & 7943             & 7987             & 7960             & 7935 \\
32000      & 1800             & 1800             & 1799             & 1794             & 1767             & 1742             & \text{-} \\
\bottomrule
\end{tabular}
\label{table:Nsample3D}
\end{table*}

\begin{table*}
\centering
\caption{The number of samples of two-dimensional packings. The numbers are presented for each number of particles $N$ and each pressure $p$.}
\sisetup{table-format=6}
\begin{tabular}{SSSSSSSSS}
\toprule
\text{$N$ / $p$} & \text{$2 \times 10^{-2}$} & \text{$10^{-2}$} & \text{$10^{-3}$} & \text{$10^{-4}$} & \text{$10^{-5}$} & \text{$10^{-6}$} & \text{$10^{-7}$} & \text{$10^{-8}$}\\
\midrule
\text{Harmonic} & & & & & & & & \\
1000             & 10945                     & 10556            & 10312            & 9628             & 9145             & 9075             & 9126             & \text{-} \\
4000             & 6013                      & 6017             & 5981             & 5863             & 5510             & 4718             & 4551             & \text{-} \\
8000             & 3709                      & 3708             & 3704             & 3664             & 3724             & 3538             & 3421             & \text{-} \\
32000            & 1330                      & 1330             & 1330             & 1329             & 1317             & 1298             & 1163             & \text{-} \\
64000            & 696                       & 696              & 696              & 695              & 694              & 648              & 426              & \text{-} \\
\midrule
\text{Hertzian} & & & & & & & \\
1000             & 7992                      & 7986             & 7946             & 7820             & 7589             & 7135             & 6964             & 5699 \\
32000            & \text{-}                  & 576              & 576              & 576              & 574              & 575              & 556              & 289 \\
\bottomrule
\end{tabular}
\label{table:Nsample2D}
\end{table*}

\section*{Conflicts of interest}
There are no conflicts to declare.

\section*{Data availability}
Data relevant to this work can be accessed at the University of Osaka Institutional Knowledge Archive~\cite{OUKA}.

\begin{acknowledgments}
KS acknowledges Yusuke Hara for useful discussions in the early stages of this work.
This work is supported by JSPS KAKENHI Grant Number 25H01519 and JST ERATO Grant Number JPMJER2401.
\end{acknowledgments}

\end{document}